\documentclass[a4paper,fleqn,usenatbib]{mnras}

\usepackage[T1]{fontenc}
\usepackage{ae,aecompl}

\usepackage{graphicx}	
\usepackage{amsmath}	
\usepackage{amssymb}	
\usepackage{deluxetable}

\title[Interpreting the time variable RM in Mrk\,421]{Interpreting the time variable RM observed in the core region of the TeV blazar Mrk\,421}

\author[R.~Lico et al.]{
R.~Lico,$^{1,2}$\thanks{E-mail: rocco.lico@unibo.it}
J.~L.~G\'omez,$^{3}$
K.~Asada$^{4}$
and A.~Fuentes$^{3}$
\\
$^{1}$Dipartimento di Fisica e Astronomia, Universit\`a di Bologna, via Gobetti 93/3, 40129 Bologna, Italy\\
$^{2}$INAF Istituto di Radioastronomia, via Gobetti 101, 40129 Bologna, Italy\\
$^{3}$Instituto de Astrof\'{\i}sica de Andaluc\'{\i}a, IAA-CSIC, Apdo. 3004, 18080 Granada, Spain\\
$^{4}$Institute of Astronomy and Astrophysics, Academia Sinica, PO Box 23-141, Taipei 10617, Taiwan
}

\date{Accepted XXX. Received YYY; in original form ZZZ}

\pubyear{2016}

\begin{document}
\label{firstpage}
\pagerange{\pageref{firstpage}--\pageref{lastpage}}
\maketitle

\begin{abstract}
In this work we interpret and discuss the time variable rotation measure (RM) found, for the first time over a 1-yr period, in the core region of a blazar. These results are based on a 1-yr, multifrequency (15, 24, and 43\,GHz) Very Long Baseline Array (VLBA) monitoring of the TeV blazar Markarian 421 (Mrk\,421). We investigate the Faraday screen properties and its location with respect to the jet emitting region. 
Given that the 43\,GHz radio core flux density and the RM time evolution suggest a similar trend, we explore the possible connection between the RM and the accretion rate. 
Among the various scenarios that we explore, the jet sheath is the most promising candidate for being the main source of Faraday rotation. During the 1-yr observing period the RM trend shows two sign reversals, which may be qualitatively interpreted within the context of the magnetic tower models. 
We invoke the presence of two nested helical magnetic fields in the relativistic jet with opposite helicities, whose relative contribution produce the observed RM values. The inner helical field has the poloidal component ($B_{\rm p}$) oriented in the observer's direction and produces a positive RM, while the outer helical field, with $B_{\rm p}$ in the opposite direction, produces a negative RM. We assume that the external helical field dominates the contribution to the observed RM, while the internal helical field dominates when a jet perturbation arises during the second observing epoch. Being the intrinsic polarization angle parallel to the jet axis, a pitch angle of the helical magnetic field $\phi\gtrsim 70^\circ$ is required.
Additional scenarios are also considered to explain the observed RM sign reversals.
\end{abstract}



\section{Introduction}
The polarized emission observed from relativistic jets in active galactic nuclei (AGNs) provides us with important information on the magnetic field configuration.
The main issue to be taken into account is that when a polarized wave propagates through a non-relativistic magnetized plasma it is affected by Faraday rotation.
As a consequence, the observed polarization angle ($\chi_{\rm obs}$) is rotated with respect to its intrinsic value ($\chi_{\rm int}$).
When the magnetized Faraday rotating medium (Faraday screen) is external to the emitting region, the effect is described by a linear relationship between $\chi_{\rm obs}$ and the observing wavelength squared ($\lambda^2$) as
\begin{equation}
\chi_\mathrm{obs}=\chi_\mathrm{int} + \mathrm{RM}\times \lambda^2,
\end{equation} 
where RM is the rotation measure \citep{Gardner1966}:
\begin{equation} \label{RM_equation}
\mathrm{RM} = 812 \int n_{\rm e} \textbf{B}_{\parallel} \cdot {\rm d}l \ \ \ [\mbox{rad \ m}^{-2}],
\end{equation}
with $n_{\rm e}$ being the electron number density in cm$^{-3}$, $\textbf{B}_{\parallel}$ the parallel to the line-of-sight, aberrated by relativistic motion, component of the magnetic field in milligauss, and d$l$ the path length in parsecs.

The determination of the RM is essential to obtain the intrinsic orientation of the polarization vectors and to properly interpret the magnetic field configuration. As shown by \citet{Zavala2004} an RM of the order of 250 rad m$^{-2}$ can produce a $25^{\circ}$ rotation of the intrinsic polarization angle at 8\,GHz. 

Polarimetric multifrequency Very Long Baseline Interferometry (VLBI) observations represent the most suitable and powerful tool for determining the RM, by measuring the electric vector position angle (EVPA) as a function of the frequency. This method allows us to investigate the structure and physics of the magnetic field in AGN jets at high angular resolution \citep[e.g.,][]{Asada2002, Zavala2001, Zavala2003, Zavala2004, Osullivan2009, Asada2010, Gomez2011, Hovatta2012, Lico2014, Gomez2016, Kravchenko2017}.

Moreover, given that the RM is also related to the thermal electron density ($n_{\rm e}$ in equation.~\ref{RM_equation}), in principle it can be used as a diagnostic tool for the accretion flow into the central black hole \citep{Bower2003, Marrone2006, Marrone2007, Kuo2014, Plambeck2014, Li2016}.

\citet{Asada2008} find a time variable RM in the jet region of the radio source 3C273 and \citet{Gomez2011} report RM sign reversals within the innermost 2 mas jet region of the radio galaxy 3C 120.
Sign reversals were also found by \citet{Osullivan2009} in the core region of three blazars within different frequency intervals and with distance from the central engine.
On the other hand, in the case of 3C 84 \citet{Plambeck2014} report a stable RM over a 2-yr period. 
In \citet[][hereafter L14]{Lico2014} we report time RM variability in the core region of the TeV blazar Markarian 421 (Mrk\,421) over a 1-yr period. This is the first time reporting a time RM variability in the core region of a blazar on a such short time scale.
Mrk\,421 is one of the best studied blazars, mainly because of its proximity \citep[$z=0.03$,][]{deVaucouleurs1991}, and represents an excellent target for the investigation of the innermost regions of relativistic jets.
The estimated black hole mass of Mrk\,421 is $M_{\rm BH}\sim 2-9 \times 10^8$ M$_{\odot}$ \citep{Sahu2016}. 
For $M_{\rm BH} = 9 \times 10^8$ M$_\odot$ we obtain a Schwarzschild radius $r_{\rm s} = 2.7 \times 10^{14}$ cm (corresponding to $8.6 \times 10^{-5}$ pc and $1.5 \times 10^{-4}$ mas).

In this paper we discuss the RM time variability reported in L14 and give an interpretation about its origin by considering the various possibilities for the location of the Faraday screen. 
We explore the possible contribution to the observed RM from both the accretion flow, by investigating the possible connection between the RM and the accretion rate, and the jet sheath, by giving a qualitative interpretation within the context of the magnetic tower models.

The paper is organized as follows: in Sect.~\ref{faraday_analysis} we describe the 43\,GHz flux density and the Faraday rotation analysis; in Sect.~\ref{discussion} we investigate and discuss the possible location of the Faraday screen and we present a short summary in Sect.~\ref{summary}.
Throughout this paper we use a $\Lambda$CDM cosmology with $h = 0.71$, $\Omega_m = 0.27$, and $\Omega_\Lambda=0.73$ \citep{Komatsu2011}.

\begin{figure}
\includegraphics[width=1.0\columnwidth]{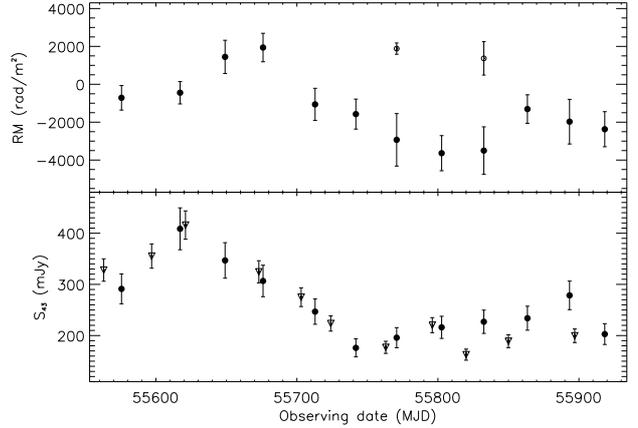} \\
\caption{Upper panel: time evolution for the core region RM values obtained from the $\lambda^2$ fits. The empty circles represent the RM values obtained without applying a $90^{\circ}$ rotation for the 15\,GHz EVPAs (see Sect.~\ref{faraday_analysis}).
Lower panel: 43\,GHz total intensity light curve. Filled circles represent data from L14, empty triangles represent data provided by the VLBA-BU-BLAZAR monitoring programme$^{\ref{BU_program}}$. Both frames of this image are adapted from Fig.~2 and Fig.~5 in L14.} 
\label{rm_vs_totint}
\end{figure}

\section{Faraday rotation analysis and 43\,GHz flux density}
\label{faraday_analysis}
Mrk\,421 was observed once per month throughout 2011 with the American Very Long Baseline Array (VLBA) at 15, 24, and 43\,GHz in total and polarized intensity (in right and left circular polarization). 
The details about the observations and calibration procedure are presented in \citet{Lico2012} for the 15 and 24\,GHz data sets and in \citet{Blasi2013} for the 43\,GHz data sets. 
The polarization calibration and analysis for the full data set are described in L14. 

The core region RM values are obtained by performing linear fits of EVPAs versus $\lambda^2$, and are reported in Table~3 in L14. The RM values vary between $(-3640 \pm 930)$ and $(+1940\pm750)$ rad\,m$^{-2}$, and in some cases are consistent with 0 within $1\sigma$ or $2\sigma$, due to the large uncertainties. 
The RM time evolution is shown in the upper panel of Fig.~\ref{rm_vs_totint}.
As described in L14, the RM values during the 7th and 9th observing epochs were obtained by using $90^{\circ}$ rotated 15\,GHz EVPAs (filled circles in the upper panel of Fig.~\ref{rm_vs_totint}), owing to optically thin/thick transitions \citep[e.g., ][]{Gabuzda2001}. 
The scenario within the context of the magnetic tower models proposed in Sect.~\ref{sec_sheat} to explain the RM trend refers to this specific EVPA configuration.
The RM values obtained without introducing any rotation for the 7th ($1880\pm 300$ rad\,m$^{-2}$) and the 9th ($1370\pm 880$ rad\,m$^{-2}$) observing epochs are represented by empty circles in the upper panel of Fig.~\ref{rm_vs_totint}.

From the linear fits of EVPAs versus $\lambda^2$ we also obtain the intrinsic polarization angle, which is roughly parallel to the jet axis ($\sim150^{\circ}$) during the entire observing period (lower panel of Fig.~5 in L14).
\addtocounter{footnote}{1}\footnotetext{\url{http://www.bu.edu/blazars/VLBAproject.html}\label{BU_program}.}
In the lower panel of Fig.~\ref{rm_vs_totint} we show the 43\,GHz light curve in which the filled circles represent data from L14, while empty triangles represent observations provided by the VLBA-BU-BLAZAR monitoring program$^{\ref{BU_program}}$. The 43\,GHz flux density ($S_{43}$) values are reported in Tables~2 and 5 in L14.
The source has an average $S_{43}$ of $\sim300$ mJy, showing an enhanced activity during the first half of 2011 (between MJD 55562 and MJD 55660). 

\section{Discussion: location of the Faraday screen}
\label{discussion}
In this section we investigate and discuss the Faraday screen properties and its location with respect to the emitting region.

Because of the relativistic mass correction, the contribution to the Faraday rotation of relativistic electrons is reduced by a factor of $\sim \gamma^2$ \citep{Jones1977, Wardle2003}, and therefore it is expected that most of the observed RM is produced by thermal electrons \citep[e.g., ][]{Celotti1998a} external to the emitting region.
The low-energy tail of the non-thermal distribution may also lead to some small {\ internal} Faraday rotation. However, given that most of the observing epochs show acceptable linear fits of EVPAs versus $\lambda^2$ (Fig.~4 in L14) and that we observe a similar trend of the fractional polarization for all of the three observing frequencies (Fig.~2 in L14), we assume that the Faraday screen is mostly external to the emitting region.

The broad line region (BLR), which extends for less than one pc and has a small ($\sim1$\%) volume covering factor \citep[e.g.,][]{Urry1995, Kaspi2000}, is usually excluded as the source of Faraday rotation.
Due to the short observed RM variability time-scale, the narrow line region (NLR), extending up to hundreds of parsecs, cannot be the primary Faraday rotation source. Moreover, there is still a large uncertainty for the NLR volume covering factor \citep{Netzer1993, Rowan-Robinson1995, Kharb2009}.
We also exclude the intracluster medium as the possible source of the observed Faraday rotation. RM values as high as the ones that we measure can originate in the intracluster medium only for sources lying in proximity of the center of cooling flow clusters, and such perfect alignment is implausible \citep{Carilli2002}. 

Two possible contributions to the observed RM remain: the accretion flow around the black hole and the sheath in the radio jet. In the following we separately treat these two scenarios.

\subsection{RM from the accretion flow}
The accretion flow region around the black hole is a possible candidate as a source of Faraday rotation.

In general, by making some simplifying assumptions on the density profile and the magnetic field in the accretion flow, in various works the observed RM values were used to constrain the mass accretion rate $\dot{M}$. 
So far, this approach was used for the accretion rate estimate for the radio sources Sgr\,A$^*$ \citep{Marrone2006}, M87 \citep[][]{Kuo2014}, and 3C\,84 \citep{Plambeck2014}.
We now try to understand if also in the case of Mrk\,421 a reliable estimate of the accretion rate could be obtained from the observed RM. 
To apply this method a roughly spherical accretion flow is required, with a power-law distribution for the radial density profile ($n \propto$ $r^{-\beta}$) and a radial, ordered and of equipartition strength magnetic field.
The density power-law index $\beta$ depends on the used accretion flow model: it varies from 3/2 for advection dominated accretion flow \citep[ADAF, ][]{Narayan1994} to 1/2 for convection dominated accretion flow \citep[CDAF, ][]{Quataert2000} models.

A sensitive parameter in the formulation of the RM as a function of the accretion rate is the inner boundary of the accretion flow region ($r_{\rm in}$), representing the distance from the black hole at which electrons become relativistic and their contribution to the RM becomes negligible.
In the case of Mrk\,421, we set $r_{\rm in}$ by using the size of the Gaussian modelfit core component of 0.04 mas \citep{Blasi2013}, which corresponds to 275 $r_{\rm s}$. By assuming an ADAF model for the accretion flow region \citep[see e.g. ][]{Celotti1998b} and by using an RM value of 2000 rad/m$^2$ we estimate an accretion rate of $\sim 2.5 \times 10^{-5}$ M$_{\odot}$/yr. 
On the other hand, if we compare the $\dot{M}$ estimated with the bolometric luminosity ($L$) by assuming a radiative efficiency of 10\% ($\dot{M}$ $\sim$ $L$/($0.1\times c^2)$), we obtain an accretion rate of $\sim 1.5 \times 10^{-2}$ M$_\odot/$yr. 
A similar discrepancy is also obtained for 3C\,84 by \citet{Plambeck2014}. As is discussed in \citet{Plambeck2014}, it could indicate that the magnetic field is weaker than the equipartition value and/or is not ordered (tangled), or the accretion flow is not spherical (possibly disc/torus like).

During the 1-yr Mrk\,421 observing campaign, RM sign reversals were observed.
An RM sign change can be ascribed only to a change in the line-of-sight magnetic field term.  
If the observed RM traces the accretion flow, it suggests that the topology of the magnetic field could be rather more complex than a radial, ordered and of equipartition strength magnetic field, and it could be attributed to the turbulence within the accretion flow.
Moreover, in the case of an RM associated with the accretion rate, an increase in the accretion rate would later lead to higher emission in the jet. 
On the contrary, the peak in $S_{43}$ occurs before the peak observed in the RM.

Therefore, in the case of the blazar Mrk\,421, all of the above mentioned arguments suggest that the accretion flow is not spherical (possibly disc/torus like structure), and the observed RM may not trace the accretion flow.

\subsection{RM from the jet sheath}
\label{sec_sheat}
Due to the peculiar blazar geometry, the outflowing plasma in the radio jet lays right in between the emitting region and the observer, so that it is reasonable to consider the jet sheath as the possible source of Faraday rotation. Thermal electrons in the jet sheath can act as a foreground Faraday screen and produce the observed RM. 

\citet{Asada2002} first proposed the jet sheath as a possible source of Faraday rotation by revealing an RM gradient across the jet of 3C 273. \citet{Zavala2004} propose a Faraday screen located in close proximity to the relativistic jet to explain the observed RM in a sample of 40 radio sources (including quasars, radio galaxies and BL Lacs) observed with VLBA between 8 and 15\,GHz.
A gradient in the RM transverse to the jet axis was revealed in various works \citep[e.g., ][]{Asada2002, Zavala2004, Osullivan2009, Gomez2008, Hovatta2012}, and suggests the presence of helical magnetic fields in relativistic jets. 

We now focus on the RM trend presented in L14 (filled circles in the upper panel of Fig.~\ref{rm_vs_totint}), with a $90^{\circ}$ 15\,GHz EVPA rotation applied during the 7th and 9th observing epochs.
Given that the source polarization properties are found to be quite stable during the 1-yr observing period, an ordered large scale magnetic field structure is expected, and we interpret the observed RM sign reversals within the context of the magnetic tower models \Citep[e.g., ][]{Lynden-Bell1996, Kato2004}. The differential rotation of the accretion disc, through the Poynting-Robertson cosmic battery effect \citep{Contopoulos1998} gives origin to nested helical fields in the relativistic jet. These models predict an inner (close to the disc symmetry axis) helical magnetic field component with the same helicity as the accretion disc rotation and an outer helical component with opposite helicity. 
In particular, the poloidal fields ($B_{\rm p}$) in the inner and outer helical components have opposite directions: $B_{\rm p}$ is parallel to the angular velocity vector ($\omega$) in the inner field component, and $B_{\rm p}$ is antiparallel to $\omega$ in the outer part \citep{Contopoulos2009}.
The net observed RM includes the contribution from both inner and outer helical field components. As shown by \citet{Mahmud2009, Mahmud2013} depending on the relative contribution of the inner and outer helical field components a change in the direction of RM gradients is expected, which can be connected with RM sign reversals.

\begin{figure}
\includegraphics[width=1.0\columnwidth]{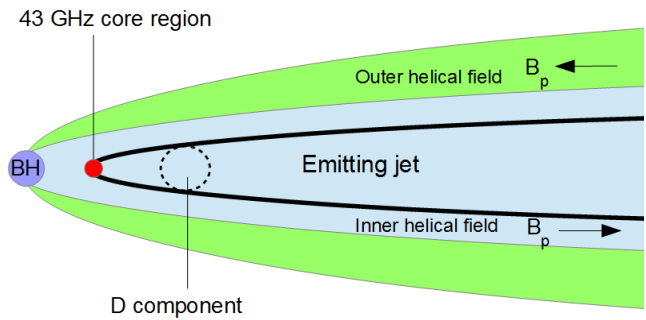}
\caption{Schematic representation of the proposed magnetic tower model. The inner helical field (blue color) extends over the emitting jet region (solid black line), up to the jet sheath region, with $B_{\rm p}$ oriented in the observer's direction and producing a positive RM. The outer helical field (green color) has $B_{\rm p}$ pointing in the opposite observer's direction and produces a negative RM.} 
\label{jet_cartoon}
\end{figure}

As shown in the cartoon in Fig.~\ref{jet_cartoon}, in the case of Mrk\,421 we assume that the inner helical field (blue color) extends over the emitting jet region (solid black line), up to the jet sheath region, with $B_{\rm p}$ oriented in the observer's direction and producing a positive RM. On the other hand, we assume that the outer helical field (green color), with $B_{\rm p}$ pointing in the opposite observer's direction, produces a negative RM and has the dominant contribution to the observed RM. 
By using the numerical model described in \cite{Gomez1995,Gomez1997}, we have performed simulations of a jet threaded by a helical magnetic field for several pitch angle ($\phi$) values and considering a viewing angle $\theta = 5^\circ$ and bulk flow Lorentz factor of $\Gamma = 1.7$ appropriate for Mrk\,421 \citep[see ][]{Lico2012}. We find that the RM sign does not depend on the direction of the magnetic field toroidal component ($B_{\rm t}$), the latter only affecting the transverse RM gradient. By estimating the Faraday-corrected polarization angle averaged across the jet width we find that a pitch angle $\phi \gtrsim 70^\circ$ ($\gtrsim 60^\circ$ in the emitting jet considering $\Gamma=1.7$) is required to explain the intrinsic polarization angle in the core region of Mrk\,421 being roughly parallel to the jet axis (at $\sim150^{\circ}$) during the entire observing period (see Sect.~\ref{faraday_analysis})

Within this scenario, an RM sign change, from negative to positive, can occur if the inner helical magnetic field in some epochs temporarily dominates the RM contribution. This behavior can be ascribed, for example, to an increase in $n_{\rm e}$ and/or in the inner magnetic field strength resulting from an increased activity in the central engine, possibly followed by the ejection of a new jet component, which produces a bow shock expanding in the neighboring regions \citep[see e.g.\,][]{Gomez1997, Fromm2016}.
Indeed, just before the observed RM sign reversals we observe an enhanced activity in the core region, both at radio and $\gamma$-ray frequencies. 
Moreover, at 43\,GHz during the second observing epoch (February 2011) a new modelfit component, labeled as D component in \citet{Blasi2013}, is observed at $\sim 0.3$ mas ($\sim 0.1$ pc) from the radio core. 
The D component could be connected with the temporary increase of the relative contribution of the inner field during the period in which a positive RM is measured.
Although this model provides a general qualitative interpretation, we note that further dedicated relativistic magnetohydrodynamical models would be required for a more quantitative description, which fall outside the scope of this paper.

Within the Poynting-Robertson cosmic battery mechanism, the direction of the accretion disc rotation univocally determines the the polarity of $B_{\rm p}$ \citep[see ][]{Contopoulos2009}.
With the specific magnetic field configuration that we propose (inner helical field $B_{\rm p}$ component positive in the observer's direction), by looking at the closely aligned to the line-of-sight jet, we are able to infer that the disc of Mrk\,421 is counter-clock-wise rotating.

On the other hand, in the case in which the 15\,GHz EVPAs during the 7th and 9th observing epochs are not rotated by $90^{\circ}$ (see Sect.~\ref{faraday_analysis}) the rapid RM changes are more difficult to explain with a magnetic tower model.
Additional scenarios to account for RM sign reversals have been investigated. 
\citet{Osullivan2009} showed how, by assuming that the Faraday rotating sheath is moderately relativistic, small changes in the velocity of the flow or slight bends of the parsec scale jet can lead to RM sign reversals in short time scales. This effect becomes relevant as long as the viewing angle is close to $1/\Gamma$. For this reason, it is not straightforward to interpret the Mrk\,421 RM sign reversals within this scenario because a very large Lorentz factor is required ($\Gamma \sim 12$) for the jet sheath, being the viewing angle $\sim5^\circ$. 
Another interesting scenario is reported in \citet{Broderick2009}, in which the authors show how RM sign reversals can arise in the transition regions between ultra-relativistic and moderately relativistic helical motion in the AGN core proximity.

The hypothesis of a Faraday screen located in the jet sheath, rather than in the accretion flow, is further supported by the fact that in Mrk\,421 the RM peak follows the peak in the flux density (Fig.~\ref{rm_vs_totint}).
This behavior is more evident if we focus on the case in which a $90^{\circ}$ rotation is applied during the 7th and 9th 15\,GHz EVPAs (filled circles in the upper panel of Fig.~\ref{rm_vs_totint}). 
To look for and quantify any possible correlation between $S_{43}$ and the RM values, we perform a discrete cross-correlation function \citep[DCF, ][]{Edelson1988} analysis, by taking into account a possible time lag.
By using a time bin of 28 d, we investigate the delay over a range of $\pm100$ d. The highest correlation value $r = 0.8$ is obtained for a 28-day time lag, at a $3\sigma$ significance level ($p$-value $\sim 0.002$), indicating that the Faraday screen is possibly located at a distance of $\sim 0.1$ pc from the emitting region.
Moreover, we note that the 43\,GHz polarized emission in the core region shows a peak during the third observing epoch, which is coincident with the increase in RM (Fig.~2 in L14), which further supports the hypothesis that the RM is produced in the jet sheath.

\section{Summary and concluding remarks}
\label{summary}
In this paper we discuss the time variable RM found for the first time over a 1-yr period in the core region of a blazar. The discussion is based on the results presented in L14, regarding a 1-yr, multifrequency (15, 24, and 43\,GHz) VLBA monitoring of the TeV blazar Mrk\,421. 
We explore the possible connection between the RM and the accretion rate, and, among the various scenarios for the identification of the Faraday screen location, we identify the jet sheath as the most suitable candidate for being the main source of Faraday rotation. 

In the specific case in which an EVPA 90$^{\circ}$ rotation is applied at 15\,GHz during the 7th and 9th observing epochs, to explain the one year RM trend, as well as the two observed RM sign reversals, we propose a magnetic tower model with two nested helical fields in the relativistic jet with opposite helicities, produced by means of the Poynting-Robertson battery effect. The inner helical field, with $B_{\rm p}$ oriented in the observer's direction, produces a positive RM and the outer helical field, with $B_{\rm p}$ in the opposite direction, produces a negative RM. We assume that the external helical field has the dominant contribution to the observed RM, while the internal field dominates when a perturbation arises during the second observing epoch. From simulations we find that, within the proposed model, the direction of $B_{\rm t}$ does not affect the RM sign and a pitch angle $\phi \gtrsim 70^\circ$ is required to reproduce an intrinsic polarization angle roughly parallel to the jet axis.
Within the assumptions of the proposed magnetic tower model, we infer that the disc of Mrk\,421 is counter-clock-wise rotating.

On the other hand, when the 15\,GHz EVPAs are not $90^{\circ}$ rotated during the 7th and 9th observing epochs, the interpretation of the RM trend is more difficult due to the rapid RM sign changes. One possibility is that, by assuming that the jet sheath is moderately relativistic, the RM sign reversals can be connected with small changes in the flow speed or slight bends of the parsec scale jet.

From our analysis it clearly emerges that the RM variability has not a straightforward interpretation. 
The Faraday rotation could be produced from different regions with different field configurations and internal Faraday rotation can also play a role and contribute to the observed RM.

\section*{Acknowledgements}
\begin{small}
We thank the anonymous referee for valuable comments and suggestions that improved the quality of this manuscript.
RL gratefully acknowledges the financial support and the kind hospitality from the Instituto de Astrof\'{\i}sica de Andalucia (IAA-CSIC) in Granada.
This work is based on observations obtained through the BG207 VLBA project, which makes use of the Swinburne University of Technology software correlator, developed as part of the Australian Major National Research Facilities Programme and operated under license \citep{Deller2011}. The National Radio Astronomy Observatory is a facility of the National Science Foundation operated under cooperative agreement by Associated Universities, Inc. We acknowledge financial support from grants PRIN-INAF-2014 and AYA2013-40825-P. This study makes use of 43 GHz VLBA data from the VLBA-BU Blazar Monitoring Program (VLBA-BU-BLAZAR;
\url{http://www.bu.edu/blazars/VLBAproject.html}), funded by NASA through the Fermi Guest Investigator Program. The VLBA is an instrument of the Long Baseline Observatory. The Long Baseline Observatory is a facility of the National Science Foundation operated by Associated Universities, Inc. \\
\end{small}



\bsp	
\label{lastpage}
\end{document}